\documentclass[letterpaper]{article} 

\RequirePackage{amsmath,amsfonts,amssymb}
\usepackage{graphicx}
\usepackage{caption}
\usepackage{subcaption}
\usepackage{multirow}
\usepackage{color}
\usepackage{lineno}
\usepackage{lscape}
\usepackage{booktabs}
\usepackage{footnote}
\usepackage[numbers]{natbib}
\title{Deep Inference of Personality Traits by Integrating Image and Word Use in Social Networks}

\author{Guillem Cucurull $^{1,}$*,  Pau Rodr\'iguez$^{1}$, V. Oguz Yazici$^{2}$, \\ Josep M. Gonfaus$^{3}$, F. Xavier Roca$^{1}$, Jordi Gonz\`alez$^{1}$ \\
	*gcucurull@cvc.uab.cat}

\begin{document}
	\maketitle
\footnotetext[1]{Computer Vision Center and Universitat Aut\`onoma de Barcelona, Edifici O, 08193 Bellaterra (Catalonia Spain)}
\footnotetext[2]{Wide Eyes Technology, C/ Galileo 303-305, 08028 Barcelona (Catalonia Spain)}
\footnotetext[3]{Visual Tagging Services, Edifici O, 08193 Bellaterra (Catalonia Spain)}
	\begin{abstract}Social media, as a major platform for communication and information exchange, is a rich repository of the opinions and sentiments of 2.3 billion users about a vast spectrum of topics. To sense the whys of certain social user's demands and cultural-driven interests, however, the knowledge embedded in the 1.8 billion pictures which are uploaded daily in public profiles has just started to be exploited since this process has been typically been text-based. Following this trend on visual-based social analysis, we present a novel methodology based on Deep Learning to build a combined image-and-text based personality trait model, trained with images posted together with words found highly correlated to specific personality traits. So the key contribution here is to explore whether OCEAN personality trait modeling can be addressed based on images, here called \emph{Mind{P}ics}, appearing with certain tags with psychological insights. We found that there is a correlation between those posted images and their accompanying texts, which can be successfully modeled using deep neural networks for personality estimation. The experimental results are consistent with previous cyber-psychology results based on texts or images. In addition, classification results on some traits show that some patterns emerge in the set of images corresponding to a specific text, in essence to those representing an abstract concept. These results open new avenues of research for further refining the proposed personality model under the supervision of psychology experts.
	\end{abstract}

\section{Introduction}
%
%
%
%

Image sharing in social networks has increased exponentially in the past years. For example, according to official Instagram stats, there are 600 million Instagrammers uploading around 100 million photos and videos per day. Although the analysis of trends, topics and brands in social networks is mostly based solely on texts, the analysis of such a vast amount of images is starting to play an important role for understanding and predicting human decision making, while becoming essential for digital marketing, and customer understanding, among others. Previous works have proven the relation between text and the personality of the authors \cite{golbeck2011predicting, iacobelli2011large}, and recent studies have also shown that some image features can be related to the personality of users in social networks \cite{cristani2013unveiling}. 

The main hypothesis of this work is that the relation between text and personality observed by researchers like Yarkoni \cite{yarkoni2010personality} translates well into a relation between images and personality when we consider the images conditioned on specific word use. In his work, Yarkoni proved that there exist words that correlate with different personality traits with statistical evidence, see Table \ref{tab:yarkoni_words}. For example, a neurotic personality trait correlates positively with negative emotion words such as 'awful' or 'terrible', whereas an extroverted person correlates positively with words reflecting social settings or experiences like 'bar', 'drinking' and 'crowd'. Considering this proven relation between text and personality, and the fact that posted images have a relation with their accompanying texts, we propose a methodology which, taking advantage of such existing text-personality correlation, exploits the relation between texts and images in social networks to determine those images most correlated with personality traits. The final aim is to use this set of images to train a personality model with similar performances than previous work based on texts or images alone. 

\begin{table*}[!t]
\renewcommand{\arraystretch}{1.3}
\caption{List of words highly related to each personality trait, extracted from \cite{yarkoni2010personality}}
\label{table:words}
\centering
\begin{tabular}{|l|c|p{10cm}|}
\hline
\textbf{Trait} & \textbf{Correl.} & \textbf{Related Words} \\
\hline
\multirow{2}{*}{O} & High & culture, films, folk, humans, literature, moon, narrative, novel, poet, poetry, sky \\ \cline{2-3}
& Low & anniversary, detest, diaper, hate, hatred, hubby, implore, loves, prayers, thankful, thanks \\
\hline
\multirow{2}{*}{C} & High & achieved, adventure, challenging, determined, discipline, persistence, recovery, routine, snack, vegetables, visit\\ \cline{2-3}
& Low & bang, bloody, boring, deny, drunk, fool, protest, soldier, stupid, swear, vain \\
\hline
\multirow{2}{*}{E} & High & bar, concert, crowd, dancing, drinking, friends, girls, grandfather, party, pool, restaurant\\ \cline{2-3}
& Low & blankets, books, cats, computer, enough, interest, knitting, lazy, minor, pages, winter \\
\hline
\multirow{2}{*}{A} & High & afternoon, beautiful, feelings, gifts, hug, joy, spring, summer, together, walked, wonderful\\ \cline{2-3}
& Low & asshole, bin, cost, drugs, excuse, harm, idiot, porn, sexual, stupid, violence \\
\hline
\multirow{2}{*}{N} & High & annoying, ashamed, awful, horrible, lazy, sick, stress, stressful, terrible, upset, worse\\ \cline{2-3}
& Low & completed, county, ground, later, mountain, oldest, poem, road, southern, sunset, thirty \\
\hline
\end{tabular}
\label{tab:yarkoni_words}
\end{table*}

In the computer vision community, the relationship between language and images has been exploited to automatically generate textual descriptions from pictures \cite{vinyals2015show,karpathy2015deep,vinyals2016show}. Indeed automatic captioning can be understood as a sampling process of words from a text distribution $t$ given an image $I$, or $p(t|I)$. In this paper we aim at the opposite: we want to determine $p(I|t)$, that means, those images mostly related with a specific word that is strongly associated to a particular personality trait. Once the set of images is defined, the potential relation between personality and images will be modeled using a state-of-the-art deep neural network. Classification results will suggest whether there is psychological content in certain images, like it has been previously observed for certain texts and image features.  

In our work, the human personality characterization called the \emph{Big Five} model of personality is considered \cite{digman1990personality,barrick1991big,goldberg1990alternative}. The Big Five model is a well-researched personality description, which has been proven to be consistent across age, gender, language and culture \cite{mccrae1992introduction,schmitt2007geographic}. In essence, this model distinguishes five main different human personality dimensions: Openness to experience (O), Conscientiousness (C), Extraversion (E), Agreeableness (A) and Neuroticism (N), hence it is often referred as \textit{OCEAN}. Personality traits are characterized in the OCEAN model by the following features:
\begin{itemize}
\item \emph{Openness}: Appreciate art and ideas, imaginative, aware of feelings. People with this trait tend to have artistic interests and have a certain level of intellectuality.
\item \emph{Conscientiousness}: Disciplined, dutiful, persistent, compulsive and perfectionist as opposite to spontaneous and impulsive. People with this trait tend to strive for something, and to be hard-workers and organized.
\item \emph{Extraversion}: Warm, assertive, action-oriented, and thrill-seeking. Individuals with high levels of extraversion tend to be friendly, sociable, cheerful and fond of being in company with other people.
\item \emph{Agreeableness}: Compassionate, cooperative, considerate. Agreeable people tend to be trusty, modest and optimistic.
\item \emph{Neuroticism}: Emotional instability, anxious, hostile, prone to depression. Neurotics tend to be frustrated, anxious and experience negative emotions.
\end{itemize}

The five personality traits have already been related to text \cite{yarkoni2010personality} and images \cite{segalin2016social} uploaded by users. Therefore, personality might be an important factor in the underlying distribution of the user's public posts in social media, and thus, it is possible to infer some degree of personality knowledge from such data. In this work we go a step beyond the works in \cite{yarkoni2010personality}, and \cite{segalin2016social}, showing that personality remains invariant to changes from the text domain to the image domain. Concretely, our contributions are: 
\begin{itemize}
\item A new framework for estimating users personality from images, called \emph{MindPics}, accessed using a refined set of the tags proposed in \cite{yarkoni2010personality}.
\item A personality inference model that uses \emph{MindPics}, as a proof that personality remains invariant across textual and visual domains.
\end{itemize}

\section{Related work}
The increasing growth and significance of the social media in our lives has attracted the attention of researchers, which use this data in order to infer about the personality, interests, and behavior of the users. Regarding personality, its inference has mainly been based on (i) text uploaded by users, and (ii) uploaded images.

\subsection{Text-based personality inference}

The relationship between language and personality has been studied extensively. As commented before, Yarkoni \textit{et al.} \cite{yarkoni2010personality} performed a large-scale analysis of personality and word use in a large sample of blogs whose authors answered questionnaires  to assess their personality. This way, by analyzing the text written by users whose personality is known, the author could investigate the relation between word use and personality. The results of the analysis concluded that the usage of some specific words are correlated with the personality of the blogs' authors.

Iacobelli \textit{et al.} \cite{iacobelli2011large} used a large corpus of blogs to perform personality classification based on the text of the blogs. They proved that both the structure of the text and the words used are relevant features to estimate the personality from text. Also, Oberlander \textit{et al.} \cite{oberlander2006whose} studied if the personality of blog-authors could be inferred from their personal posts. To do so, the personality profile of 71 different users was collected through a questionnaire. By using word n-grams as features and SVM classifiers they successfully managed to infer the personality of the authors of the post. 

Besides blogs, where long texts are common, personality studies based on language have also focused in much shorter texts such as the shared in the social network \textit{Twitter}, where the text shared by users is limited to a maximum of 140 characters. In a similar way, Golbeck \textit{et al.} \cite{golbeck2011predicting} showed that the personality of users from Twitter could be estimated from their \textit{tweets} taking into account also other information as the number of followers, mentions or words per \textit{tweet}. 

\subsection{Image-based personality inference}

An early attempt to model personality from images was presented in Steele \textit{et al.} \cite{steele2009your}. This work explored the characteristics of profile pictures and the relationship with the impression that other users had of the owners of these pictures. Their findings show that users better agreed with the targets’ self-reported personalities when the profile picture was a photo of a human, taken outdoors instead of indoors, and the face appears smiling. 

Cristiani \textit{et al.} \cite{cristani2013unveiling} proved that there are visual patterns that correlate with the personality traits of 300 Flickr users and thus, that personality traits of those users could be inferred from the images they tagged as favorite. To do so, two categories of visual features are extracted from the images: \textit{"Aesthetic"} and \textit{"Content"} features. Aesthetic features focus on aesthetic information of the images, aiming to represent the user preferences. This type of features include the amount of certain colors, the number of edges in the image, the entropy and the level of detail among others. Content features focus on describing what appears in the image by counting the number of faces and identifying the different objects using the GIST descriptor \cite{oliva2001modeling}.

Guntuku \textit{et al.} \cite{guntuku2015personality} improved the low level features used in previous work by changing the usual Features-to-Personality (F2P) approach to a two step approach: Features-to-Answers (F2A) + Answers-to-Personality (A2P). Instead of building a model that directly maps features extracted from an image to a personality, with this approach the features are first mapped to the answers of the questionnaire BFI-10 for personality assessment \cite{rammstedt2007measuring}. Then, the answers are mapped to a personality. Besides this two-step approach, they also add new semantic features to extract from the images, like \textit{Black \& White vs. Color image}, \textit{Gender identification} and \textit{Scene recognition}.

Later, Segalin \textit{et al.} \cite{segalin2016pictures} proposed a new set of features that better encode the information of the image used to infer the personality of the user who favourited it. They proposed to describe each image with 82 different features, divided in four major categories: \textit{Color}, \textit{Composition}, \textit{Textural Properties} and \textit{Faces}. These groups of features are similar to the ones proposed by \cite{machajdik2010affective}, but excluding the \textit{content} group and using instead the number of faces as a feature of the image. The reason behind is that faces are very common in the images and that the human brain is specifically wired to perform accurate face detection and recognition. Their method proved to be suitable to map an image to a personality trait, but it worked better for attributed personality traits rather than self-assessed personality.

Since a Convolutional Neural Network (CNN) won the Imagenet competition in 2012 \cite{krizhevsky2012imagenet} the computer vision field has moved from designing the hand-made image features to learn them in an end-to-end deep learning model. Likewise, the feasibility of deep learning to automatically learn features that are good to estimate personality traits from pictures have been already proven by the same work of Segalin \textit{et al.} \cite{segalin2016social}. In their work, they presented the dataset \textit{PsychoFlickr}, which consists of a collection of images favourited by 300 users from the site \textit{Flickr.com}, each user tagging 200 images as favorite, adding up to a total of 60,000 images. Additionally, the Big Five traits personality profile of each user are provided. There are two different versions of the personality profile for each user, one collected through a self-assessment questionnaire answered by the user, and one attributed by a group of 12 assessors who had evaluated the image set of the user. Subsequently, the authors fine tune a CNN pre-trained on the large dataset of object classification Imagenet \cite{imagenet_cvpr09} to capture the aesthetic attributes of the images in order to be able to estimate the personality traits associated with those images. For each of the Big Five traits they trained a CNN model with a binary classifier. Then, each different CNN estimates if the images are “high” or “low” for the trait the model has been trained for.
\\ The study of the personality conveyed by images has not only been used to infer the personality of users, but also to analyze how brands express and shape their identity through social networks. Ginsberg \emph{et al.} \cite{ginsberg2015instabranding} analyzed the pictures posted in Instagram by the leading food brands and classified them into different categories: product, person and product, people and product, humor and product, world events, recipes, campaign with no products, user-generated, celebrity, and video. Then, the analysis of the types of images posted by the brands was used to interpret the identity of each brand along five dimensions of personality: sincerity, excitement, competence, sophistication, and ruggedness.

\begin{table*}[!t]
\centering
\caption{Classification accuracies (\%) reported in the literature using texts or images}
\label{tab:tab_soa}
\begin{tabular}{|l||c|c||c|c|}
\hline
 & \multicolumn{2}{c||}{Word use only}  & \multicolumn{2}{c|}{Image use only}  \\ \hline
 &  Golbeck \cite{golbeck2011predicting} & Iacobelli \cite{iacobelli2011large} & Segalin \cite{segalin2016social} & Guntuku \cite{guntuku2015personality} \\ \hline \hline
O           & 75.50 & 84.36  & 61.00  & 66.10          \\ \hline
C  & 61.70 & 79.18  & 67.00  & 70.50           \\ \hline
E       & 58.60  & 71.68  & 65.00 & 69.70          \\ \hline
A       & 69.70 & 78.31   & 64.00 & 72.30         \\ \hline
N         & 42.80 & 70.51  & 69.00  & 61.50         \\ \hline \hline
\textbf{Avg}    & 61.66 & 76.80  & 65.20  & 68.02         \\ \hline
\end{tabular}
\end{table*}

\subsection{Integrating text and image for personality inference}

Table \ref{tab:tab_soa} contains the recognition accuracy of each of the OCEAN traits for 5 different methods based on text or images. In this paper, we want to determine whether the correlations found between personality and texts or images separately, also holds when combining image and word use.

Indeed all previous visual-based approaches take advantage of the many ways in which users interact with images in social networks, such as posting an image, liking it or commenting on it. Specifically, most of the works described above consist on assessing the personality of users based on the images they have liked. For example, the main difference between the \cite{segalin2016social} and our approach is that in such paper, the personality is inferred based on which images have been tagged as \textit{favorite}, thus becoming a study on the relation between aesthetic preferences and personality. In contrast, in our case, we explore directly the images shared by users based on accompanying texts strongly related to personality traits, so here the relationship between images and personality arises from the mere act of posting a picture in a social network as a process of communication with others. The whole procedure is detailed next.

\section{Methodology}

As proven by Yarkoni \textit{et al.} \cite{yarkoni2010personality} there exists a relationship between the personality of people and the language they use. In other words, the language that we use can reveal our personality traits, so there is statistical evidence that the use of specific words correlate with the personality of online users. Based on that, we design a set of images $S$ conditioned by those words most related to specific personality traits. This can be seen as sampling images from a distribution of images $I$ conditioned on text $t$, or $S \sim p(I|t)$.

For each trait of the Big Five personality traits defined before we have selected the list of words that correlate most positively or negatively with the trait, as suggested by \cite{yarkoni2010personality}. The positively correlated words have been used to identify those images most associated to the strong presence of a trait, and the negatively correlated words are used to determine those images most associated to its absence.

From this set of images $S$ we can train a deep learning model that learns to extract a personality representation from a picture, and use it to automatically infer the personality that the picture conveys.

\subsection{Finding \emph{Mind{P}ics} in Social Networks}

Based on the aforementioned relationship between text and personality, the proposed personality model is built considering a large quantity of images, called \emph{Mindpics}, tagged with specific personality-related words. These words are the most correlated ones with each personality trait, as presented in \cite{yarkoni2010personality}. In Table \ref{table:words}, we showed the words mostly related to each personality trait, which have been also used to identify that set of images for each trait used for training the neural network. So each image related to a tag will correspond to one of the five personality traits, and within the trait it will represent the \textit{high} or the \textit{low} presence of such trait.

As the aim of this paper is to evaluate whether there is any of the author's personality information embedded in real world images, we have considered images posted in Instagram \cite{ginsberg2015instabranding,hu2014we,souza2015dawn} . In this social network the users take and share images by posting them together with words or \emph{hashatgs}. To build the \emph{Mindpics} dataset, we first crawl a large collection of publicly-shared photos using the Instagram API.

The reason of choosing Instagram as the source of selecting the images for training the Neural Network is threefold: (i) images are the main content of Instagram instead of text. Contrary to more text-based social networks such as Twitter, most of the content and information shared by the Instagram user is conveyed in the image, so it is reasonable to think that such images embed personality, up to some extent. (ii) The fact that the images can be accompanied by text allows to easily identify those images that appear together with specific words. (iii) By considering in our experiments the public pictures posted by hundreds of users, these users are not aware that they are being part of a psychological experiment.

\begin{figure*}[!t]
\centering
\includegraphics[width=0.6\linewidth]{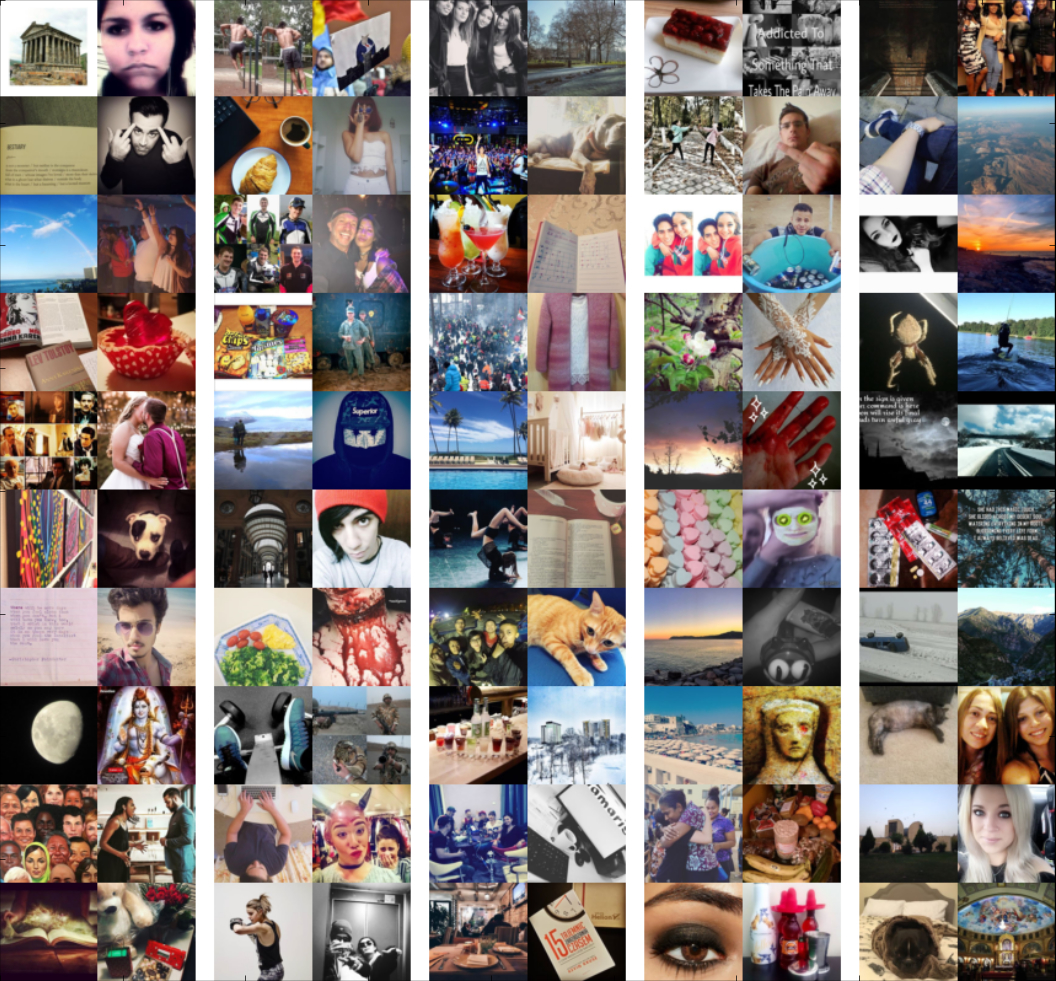}
\caption{\textbf{\emph{Mind{P}ics} Samples.} Each column contains 10 \emph{Mind{P}ics} of each personality trait. From left to right: High openness, Low openness ; High conscientiousness, Low conscientiousness ; High extraversion, Low extraversion ; High agreeableness, Low agreeableness ; High neuroticism, Low neuroticism.}
\label{fig:all_ocean}
\end{figure*}

Instagram images are very interesting, because there are no boundaries in the kind of pictures that users use to communicate with their followers. Thus, different kinds of users will post different kinds of pictures, as proven by Hu \textit{et al.} \cite{hu2014we}. In their work, they show that the pictures posted in Instagram can be classified into eight main categories, and the users can be divided in five different groups, depending on what kind of pictures they post. These eight main picture categories are: friends, food, gadget, captioned pictures, pets, activity, selfies, and fashion. The difference on the type of images posted can be influenced by the city of the users, \cite{hochman2012visualizing} or their age \cite{jang2015generation}. 

In order to determine the \emph{MindPics} set of images, we used the words from Table \ref{table:words} to query images. For each personality trait, 22 words were used, 11 for each component, and about 1,100 images were selected for each word. The total number of \emph{MindPics} images used for training the personality model is $121,000$. Because we used the same number of words per trait and the same number of images per word, the number of training images results balanced within the 5 personality traits and also within the used tags, thus each trait being trained from around $24,000$ images.

In Figure \ref{fig:all_ocean} we show some 10 random samples of each personality trait obtained with the procedure described above. From left to right the classes are: high Openness, low Openness, high Conscientiousness, low Conscientiousness, high Extraversion, low Extraversion, high Agreeableness, low Agreeableness, high Neuroticism, low Neuroticism. By observing these pictures we can see how despite the challenging variability of the data we can identify that some visual patterns emerge, like pictures containing more people and crowds for \textit{High Extraversion} than for \textit{Low Extraversion}, or that \textit{Low Neuroticism} correlates well with images depicting more nature and landscape patterns than in \textit{High Neuroticism}.

As it can be seen, images of the same class present a lot of variability, see for example how \textit{High Openness} contains images of different objects like books, faces, text, drawings, and landscapes among others. Also, we can see how images of people appear in several classes. This huge variability is a very challenging problem, because it is harder to recognize unique patterns in the images of each class.

Summarizing, despite the huge intra-class and inter-class variabilities of the images associated to each of the personality traits, we can show that there are consistence between the images of the same class, whose hashtags are related to the words suggested by \cite{yarkoni2010personality}.

\subsection{Building the Personality model}

Once we have defined the procedure to determine which set of images $S$ is most related to each personality trait, we next describe how we can model this relationship between \emph{Mind{P}ics} and personality. 

In this work we have used a neural network model that maps an input image to a desired output by learning a set of parameters that produce a good representation of the input. This model is hierarchical, \emph{i.e.} it consists of several layers of feature detectors that build a hierarchical representation of the input, from local edge features, to abstract concepts. The final layer consists on a linear classifier, which projects the last layer features into the label space. Let $x$ be an input image and $f(x;\theta)$ a parametric function that maps this input to an output, where $\theta$ are the parameters. Because a neural network model is a hierarchical combination of computation layers, the output is a composition of non-linear functions: 
\begin{equation}
f(x;\theta) = f^N(f^{N-1}(...f^2(f^1(x;\theta_1);\theta_2);\theta_{N-1});\theta_N)
\end{equation}
where $N$ is the number of layers in the model, and each computation layer corresponds to a non-linear function $f^n$ \cite{Goodfellow-et-al-2016} with its own parameters $\theta_n$. We find $\theta$ by empirical risk minimization:
\begin{equation}
\theta = \arg\min_{\theta} \mathcal{L}(y, \hat{y}),
\end{equation}
where $\mathcal{L}$ is the cross-entropy loss function, $y$ is the output of the model defined as $y = f(x;\theta)$ and $\hat{y}$ is the correct output for input $x$. This function is minimized iteratively by means of Stochastic Gradient Descent (SGD), and the process of minimizing the loss function over a set of images $S$ is referred as \emph{training} the model.
There are different types of Deep Learning architectures, in this work we use a Convolutional Neural Network (CNN) model \cite{lecun1998gradient}, \cite{krizhevsky2012imagenet}, since CNNs are especially suited for 2D data. After training, the output of the CNN for an image is a vector of scores for each of the personality traits.

For predicting the \emph{high} and \emph{low} scores for each of the personality traits, we propose two different strategies: one model for each trait, and a all-in-one model.

\subsubsection{One model for each trait}
Having one model for each trait consists on training five CNNs, using five different sets of parameters $\theta_i, i\in [1,5]$, each one to predict one of the Big Five personality traits. Namely, each CNN receives an image as input and produces a vector $o$ of two dimensions as output. This vector $o$ is used as input to the softmax function, that produces $p$, a score vector indicating the probability of an individual personality trait to be inferred from the image.
For an output vector $ o $ of two components the softmax function for each component $i$ is defined:

\begin{equation} \label{eqn:softmax}
p_i(o_i)=\frac{e^{o_i}}{\sum_{j=1}^{2} e^{o_j}}
\end{equation}

Note that $ \sum p = 1 $. Because the task to solve is a classification task, the Multinomial Logistic Loss is used as the loss function to minimize. The loss for one sample is defined as:

\begin{equation} \label{eqn:mlogloss}
\mathcal{L} = -log(p_{l}),
\end{equation}

where $p_{l}$ is the probability assigned by the CNN model of the input belonging to the true class $l$.

Note that, although training one model for each trait might seem conceptually easier, it increases the computations by a factor of five, and does not benefit from feature reusing, which is important for the model to generalize. To address these shortcomings, we propose the all-in-one model.

\subsubsection{All-in-one model}
\begin{figure}[!t]
\centering
\includegraphics[width=0.8\linewidth]{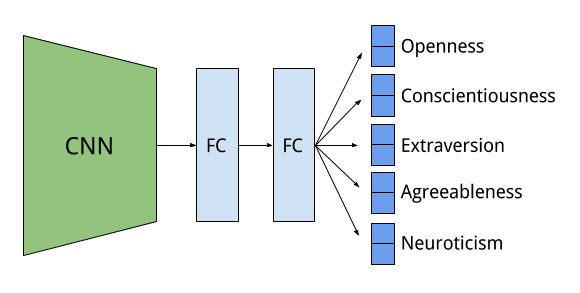}
\caption{\textbf{All-in-one-model.} Scheme of the all-in-one model that jointly trains 5 different classifiers that share parameters, one for each personality trait.}
\label{fig:all_in_one}
\end{figure}

We propose to use the same CNN for all five traits, but using five different classifiers on the last layer, one for each of the Big Five traits. Each output layer is independent of the others, and consists on a binary classifier like the described before, which has its own loss function. In Figure \ref{fig:all_in_one} it is depicted an scheme of this model.

The loss used for each of the 5 independent binary classifiers is the Multinomial Logistic Loss from Equation \ref{eqn:mlogloss} as in the single-classifier model, with one minor modification. For the multi-classifier setup we must consider that we only want to backpropagate errors made by the classifier responsible for the ground truth label. For instance, if the true label of an image is \textit{"high Openness"} the only error backpropagated should be the error produced by the \textit{"Openness"} classifier. Hence, we modify Equation  \ref{eqn:mlogloss} so that the loss for each classifier is zero if the classifier does not have to be considered: 

\begin{equation} \label{eqn:mlogloss_multi}
\mathcal{L}_c = \begin{cases}
-log(p_{l}) &\text{correct classifier}\\    
0 &\text{ignored classifier}
\end{cases}
\end{equation}

This equation defines the loss function computed for one of the independent classifiers $c$. Each classifier is independent from the others, so each classifier will compute its own loss function $\mathcal{L}_c$ and backpropagate the error. The reason for Equation \ref{eqn:mlogloss_multi} is that by setting the loss to 0 for the classifiers that do not have to be considered for the input image, the gradients backpropagated to the previous layers by those classifiers will be also 0. Hence, the only classifier that will take part in updating the network parameters is the desired one.

\section{Experiments}
In this section we describe the different models that we have tested to infer personality traits from images and explain the details of the training process. We finish the section by exposing the quantitative results of our approach and with a qualitative analysis of the model trained for personality inference.

\textbf{CNN models.}
There are some common CNN architectures \cite{krizhevsky2012imagenet, simonyan2014very, szegedy2015going, he2016deep} that are well established for computer vision tasks. In our experiments, we have used two of these CNN models. The first model is the same as proposed by Krizhevsky \textit{et al.} \cite{krizhevsky2012imagenet} with the only difference that we replace the output layer of 1000 dimensions by the two configurations explained before. This architecture receives an input image of $227\times227$ pixels which is ran through five consecutive convolutional layers. Then, the output of these convolutional layers is ran through two Fully Connected layers before feeding the output layer. From now on this model will be referred as \textit{Alexnet} for brevity. The other CNN model we have used is a Residual Network presented in \cite{he2016deep}, abbreviated to \textit{ResNet}. In this case we also change only the output layer to adequate the architecture to our task. The main difference between these two models is that the latter uses residual connections between the network layers, allowing for an easier optimization process of deep networks, and thus allowing to increase their depth. Another difference is that the ResNet architecture uses Batch Normalization \cite{ioffe2015batch} after every convolutional layer. This technique reduces the internal covariate shift in the distribution of the model activations by normalizing the inputs of each layer.  In the experiments we use a ResNet of 50 layers, whereas the Alexnet model consists of 8 layers. For both models we test the two output configurations explained previously.

\textbf{Finetunning.}
We also test two different options for initializing the network's weights: the first option is to train the model from scratch starting from random weights and the other option is the finetunning approach, which consists in initializing the network with the weights learned for another task. In this case, we initialized the network with the weights of a model trained on ImageNet \cite{russakovsky2015imagenet}. The finetunning approach has been proved \cite{oquab2014learning} to be useful to train neural networks in small datasets with superior performance. The idea behind this approach is that the network first learns how to extract good visual features in a large dataset and then uses this features to learn a classifier in a smaller dataset.

\textbf{Training setup.}
To train the models, we used the deep learning framework Caffe \cite{jia2014caffe} and a GPU NVIDIA GTX 770 with 4GB of memory. The optimization algorithm used to train the network is Stochastic Gradient Descent with momentum of 0.9. For the Alexnet model we use a batch size of 128, whereas for the ResNet network we set the batch size to 32. The learning rate is set to $0.01$ when training a network from scratch and to $0.001$ when finetunning, increasing the learning rate by a factor of $10$ at the new layers. Dropout \cite{srivastava2014dropout} is used in the Fully Connected layers with a probability $p=0.5$. $L_2$ regularization of the weights is also used with a factor of $0.0005$. Additionally, during the training stage we apply data augmentation to the input images. Namely, we randomly apply horizontal mirroring to the images and crop a random patch of $224x224$ pixels of the original $256x256$ images. The cropped patch is used as the input to the network. The only pre-processing of the images is mean subtraction. The mean image to be subtracted is computed on the training set. A random split of the Mindpics dataset is used to divide the images in non-overlapping training and testing sets, $80 \%$ of the images are used for training and $20\%$ for testing. 

\begin{table*}[!t]
\centering
\caption{Classification accuracies (\%) for each personality trait on the Mindpics dataset}
\label{tab:mindpics_results}
\begin{tabular}{c c c c c c | c}
\toprule
\textbf{Model}  & \textbf{O}  & \textbf{C}  & \textbf{E}  & \textbf{A}  & \textbf{N} & \textbf{Avg} \\
\midrule
Alexnet independent & 61.6 & 62.9 & 63.4 & 62.5 & 64.0 & 62.9 \\
Alexnet all-in-one & 62.3 & 63.5 & 63.6 & 62.1 & 65.4 & 63.4 \\
\midrule
Alexnet independent & 67.8 & 68.3 & 73.5 & 67.4 & 69.1 & 69.2 \\
Alexnet all-in-one & 66.9 & 69.2 & 73.6 &  67.8 & 69.4 & 69.4 \\ 
ResNet independent & 69.5 & \textbf{72.8} & 76.9 & 69.1 & \textbf{70.3} & 71.7 \\
ResNet all-in-one & \textbf{69.8} & 72.4 & \textbf{77.7} & \textbf{69.8} & 69.6 & \textbf{71.9} \\
\bottomrule
\end{tabular}
\end{table*}

\begin{figure*}[!t]
    \centering
    \begin{subfigure}[b]{0.49\textwidth}
        \includegraphics[width=\textwidth]{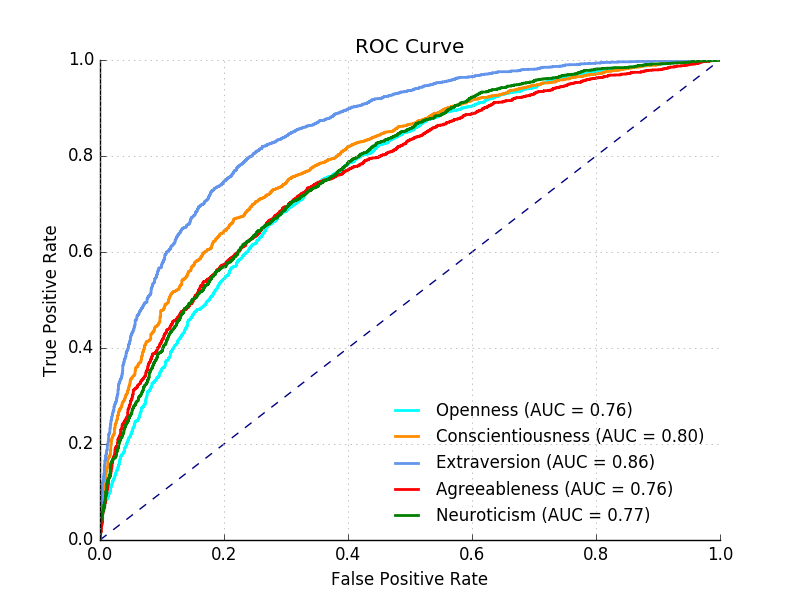}
        \caption{ROC curves for all traits}
        \label{fig:roc_curve}
    \end{subfigure}
    \begin{subfigure}[b]{0.49\textwidth}
        \includegraphics[width=\textwidth]{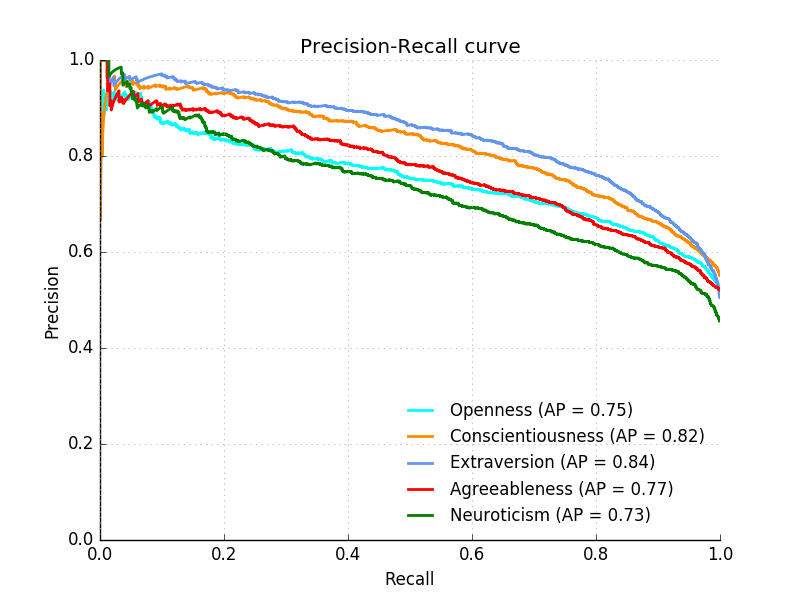}
        \caption{PR curves for all traits}
        \label{fig:pr_curve}
    \end{subfigure}
    \caption{Receiver operating characteristic and Precision-recall curves for all the personality traits. The ROC plot also shows the Area under the curve for each trait, and the PR plot shows the Average precision for each trait. }
    \label{fig:curves}
\end{figure*}


\subsection{Quantitative Results}
In Table \ref{tab:mindpics_results} the accuracies on personality recognition for each trait obtained by the different models and configurations we tested are shown. The first two rows show the performance of the Alexnet model trained from scratch with random weights initialization. The first row shows the results obtained by training an independent model for each trait, and the second row the results obtained by training just one model that shares most of the computation and that has an independent output layer for each trait. As it can be seen in the table, the \textit{all-in-one} model obtains better accuracies in all the traits except for Agreeableness, and the average accuracy is 0.5 \% better compared to training five independent models. 

The rest of the results correspond to the finetunning of Alexnet and ResNet models starting from the weights learned on the ImageNet dataset. The results show that pre-training the network in a larger dataset in order to learn to extract better features increases performance in all the personality traits, increasing the average accuracy by $6\%$. It can also be seen that the \textit{all-in-one} model performs better in this scenario too, both for Alexnet and ResNet neural networks. In these cases however, the increase of performance is smaller, the average accuracy increases by $0.2 \%$ in both cases. 

The \textit{all-in-one} configuration learns better features because the personality classifiers share the feature extraction layers and each classifier contributes to the learning, whereas a model that only has one classifier does not see as many different images so it learns worse features. However, when pre-training on the Imagenet, both networks start already with good feature extractors, so this effect is not as important as when the networks are trained from scratch. Besides the increase in accuracy performance, the \textit{all-in-one} network is also much more efficient, because it shares most of the image processing steps for all traits, thus reduces the amount of computation by a factor of five.

Finally, the results also show that the ResNet network is significaly better than the shallower network Alexnet, achieving up to $2.5\%$ more in average accuracy and consistently increasing the performance for all traits. This increase in performance can be explained from the increase in depth of the model, which allow the models to learn better representations.
To further explore the performance results, in Figure \ref{fig:curves} we show the Receiver Operating Characteristics (ROC) and Precision-recall (PR) curves for each trait obtained by best model, along with the area under the ROC curve (AUC) and Average Precision (AP). The predictions used to plot the curves are obtained from the ResNet all-in-one CNN. As it can be seen, the Extraversion trait obtains the best scores, followed by Conscientiousness. The dashed line in the ROC curves figure represents a hypothetical random classifier with and AUC of 0.5. As it can be seen, the ROC curves for each personality trait are far from being random, indicating that the relation between text and images exists and that the model is able to learn a mapping from image to personality.

\subsection{Qualitative Results}

In this section we analyze the qualitative results obtained with our deep personality model based on \emph{Mind{P}ics}. As previously seen, the best models are those pre-tained in ImageNet, which learn good features in the pre-training process, and the \textit{all-in-one} models that learn to extract better features by sharing the feature extractor layers between different personality classifiers. The relationship between the quality of the extracted features and the final classification performance has already been explored in the literature. For instance, \cite{guntuku2015personality} showed that for personality recognition improving the hand-crafted features extracted from images lead to an improvement of the results obtained with more basic features \cite{cristani2013unveiling}. 

Therefore, we can assume that the increase in performance of the models that get better results is caused because the network is able to extract a more suitable representation of the image in the feature space, a representation that allows the classifier to better discriminate the images. To get a better insight of what the new deep features are detecting, we visualize and analyze the images that maximally activate the output of the network for each personality trait.

Namely, in order to know which images better represent each personality trait, we  find those pictures that maximally activate a specific output of our model. Similarly to \cite{girshick2014rich}, we input all the images to our model, inspect the activation values of a specific neuron, and look for the images that produce these maximum activations. 

\begin{figure*}[!t]
    \centering
    \begin{subfigure}[b]{0.32\textwidth}
        \includegraphics[width=\textwidth]{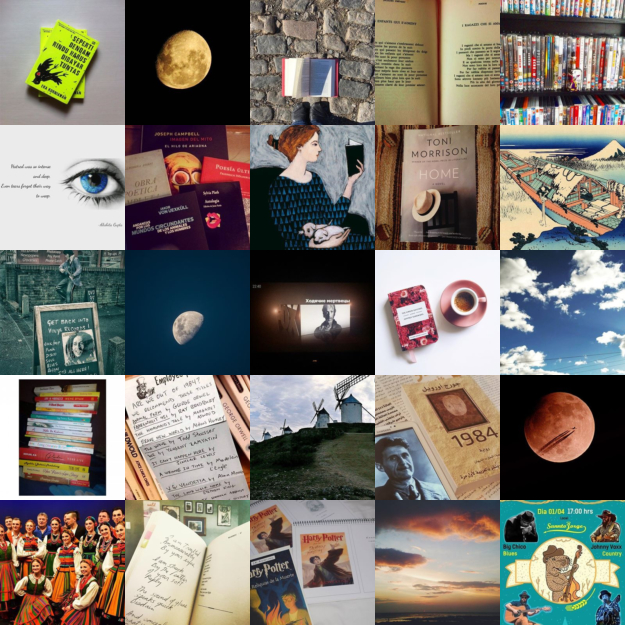}
        \caption{High Openness}
        \label{fig:max_high_o}
    \end{subfigure}
    \hspace{0cm}
    \begin{subfigure}[b]{0.32\textwidth}
        \includegraphics[width=\textwidth]{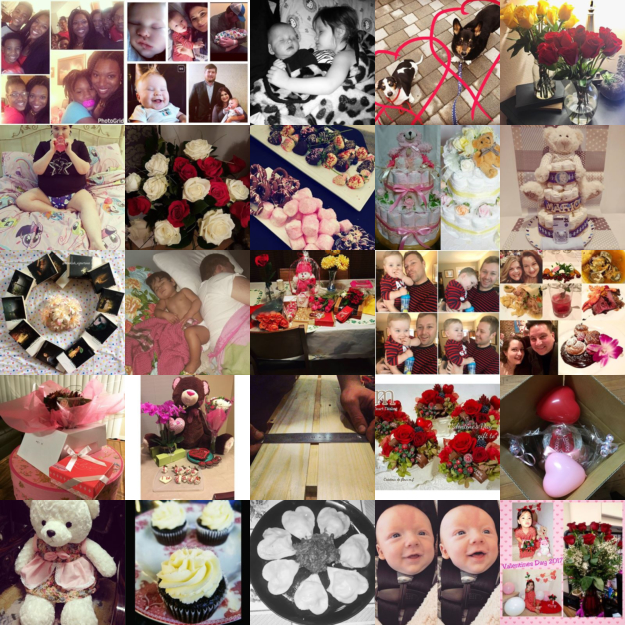}
        \caption{Low Openness}
        \label{fig:max_low_o}
    \end{subfigure}
    \begin{subfigure}[b]{0.32\textwidth}
        \includegraphics[width=\textwidth]{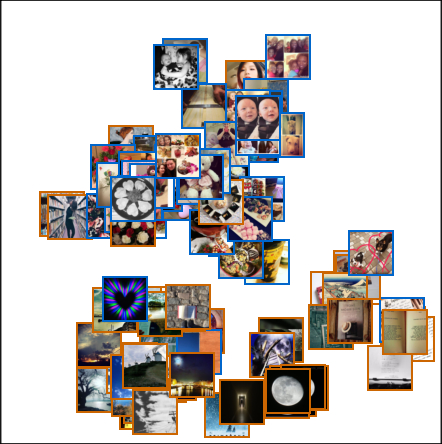}
        \caption{t-SNE visualization}
        \label{fig:tsne_open}
    \end{subfigure}
    \caption{\emph{Mind{P}ics} that maximally activate Openness.}
    \label{fig:max_openness}
\end{figure*}

\begin{figure*}[!t]
    \centering
    \begin{subfigure}[b]{0.32\textwidth}
        \includegraphics[width=\textwidth]{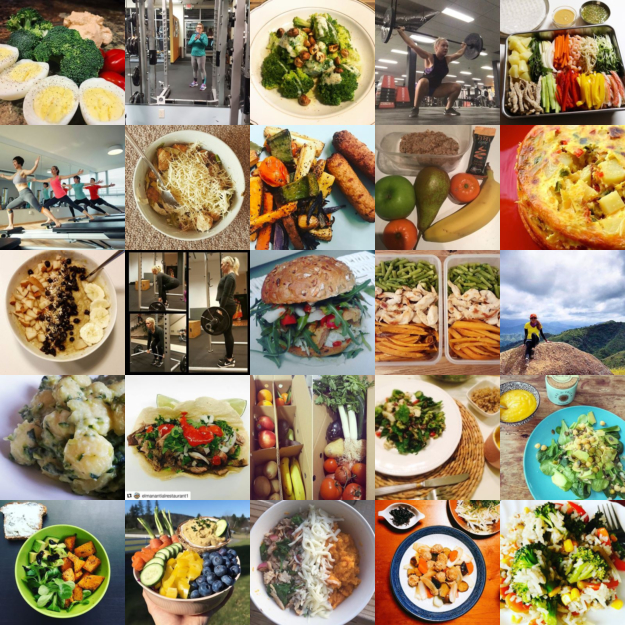}
        \caption{High Conscientiousness}
        \label{fig:max_high_c}
    \end{subfigure}
        \hspace{0cm}
    \begin{subfigure}[b]{0.32\textwidth}
        \includegraphics[width=\textwidth]{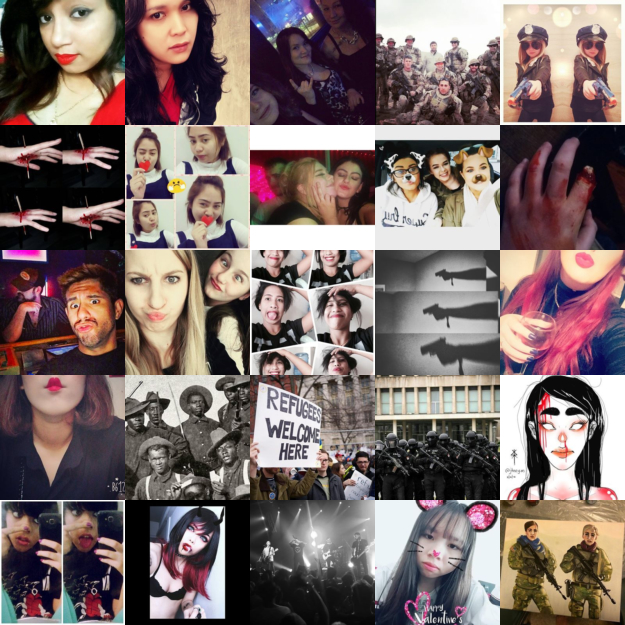}
        \caption{Low Conscientiousness}
        \label{fig:max_low_c}
    \end{subfigure}
    \begin{subfigure}[b]{0.32\textwidth}
        \includegraphics[width=\textwidth]{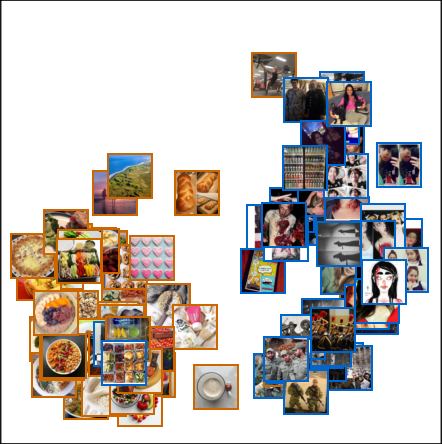}
        \caption{t-SNE visualization}
        \label{fig:tsne_con}
    \end{subfigure}
    \caption{\emph{Mind{P}ics} that maximally activate the Conscientiousness.}
    \label{fig:max_cont}
\end{figure*}

\begin{figure*}[!t]
    \centering
    \begin{subfigure}[b]{0.32\textwidth}
        \includegraphics[width=\textwidth]{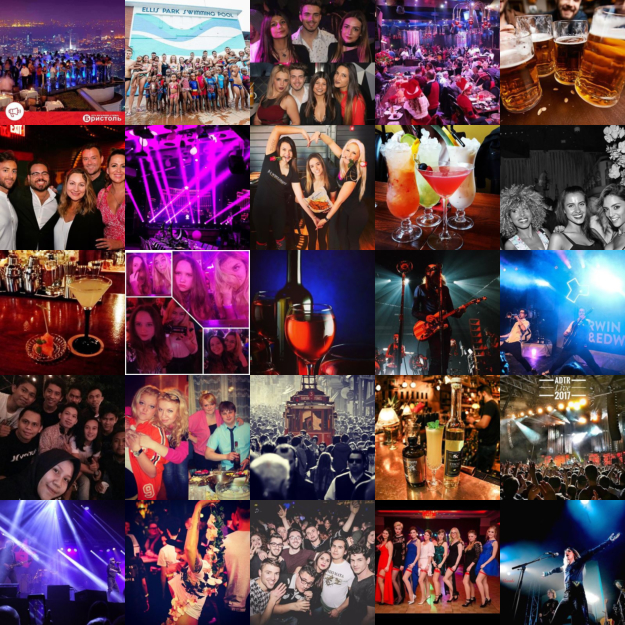}
        \caption{High Extraversion}
        \label{fig:max_high_e}
    \end{subfigure}
        \hspace{0cm}
    \begin{subfigure}[b]{0.32\textwidth}
        \includegraphics[width=\textwidth]{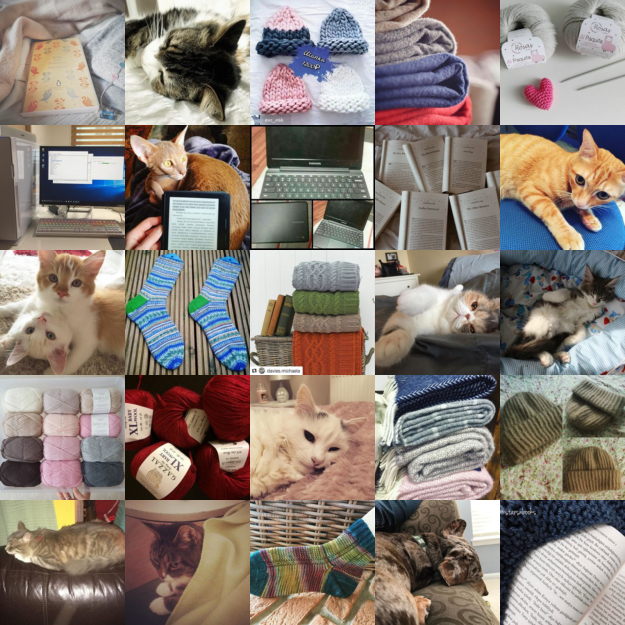}
        \caption{Low Extraversion}
        \label{fig:max_low_e}
    \end{subfigure}
     \begin{subfigure}[b]{0.32\textwidth}
        \includegraphics[width=\textwidth]{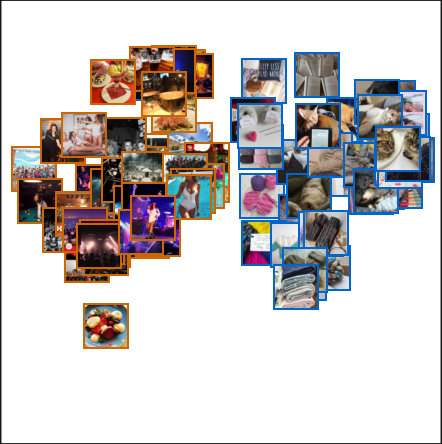}
        \caption{t-SNE visualization}
        \label{fig:tsne_extr}
    \end{subfigure}
    \caption{\emph{Mind{P}ics} that maximally activate the Extraversion.}
    \label{fig:max_extraversion}
\end{figure*}

\begin{figure*}[!t]
    \centering
    \begin{subfigure}[b]{0.32\textwidth}
        \includegraphics[width=\textwidth]{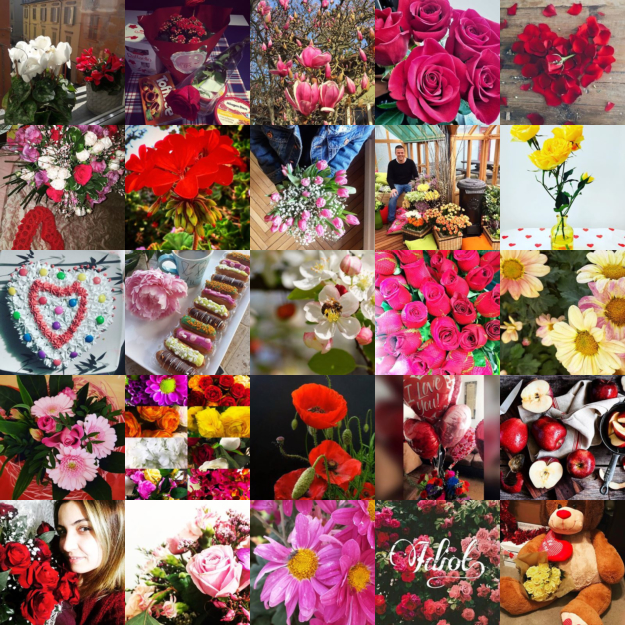}
        \caption{High Agreeableness}
        \label{fig:max_high_a}
    \end{subfigure}
        \hspace{0cm}
    \begin{subfigure}[b]{0.32\textwidth}
        \includegraphics[width=\textwidth]{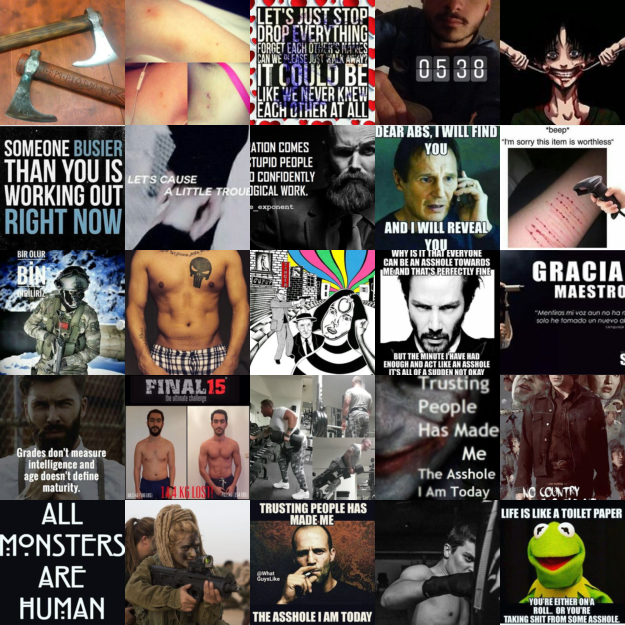}
        \caption{Low Agreeableness}
        \label{fig:max_low_a}
    \end{subfigure}
     \begin{subfigure}[b]{0.32\textwidth}
        \includegraphics[width=\textwidth]{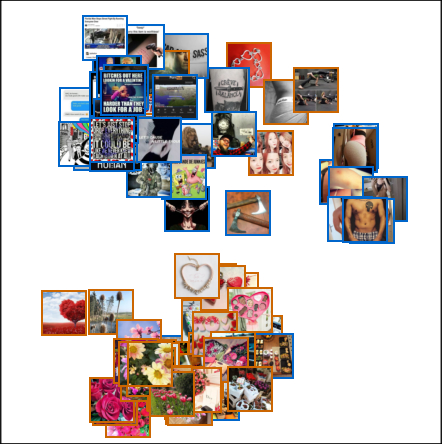}
        \caption{t-SNE visualization}
        \label{fig:tsne_agre}
    \end{subfigure}
    \caption{\emph{Mind{P}ics} that maximally activate the Agreeableness.}
    \label{fig:max_agre}
\end{figure*}

\begin{figure*}[!t]
    \centering
    \begin{subfigure}[b]{0.32\textwidth}
        \includegraphics[width=\textwidth]{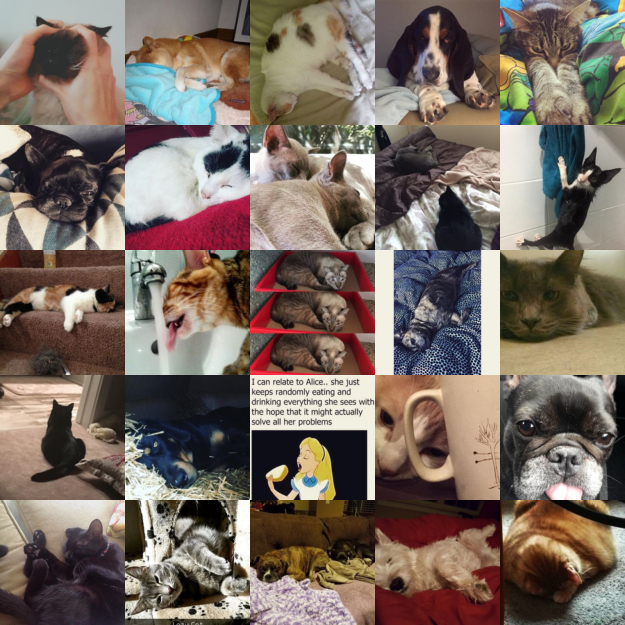}
        \caption{High Neuroticism}
        \label{fig:max_high_n}
    \end{subfigure}
    \hspace{0cm}
    \begin{subfigure}[b]{0.32\textwidth}
        \includegraphics[width=\textwidth]{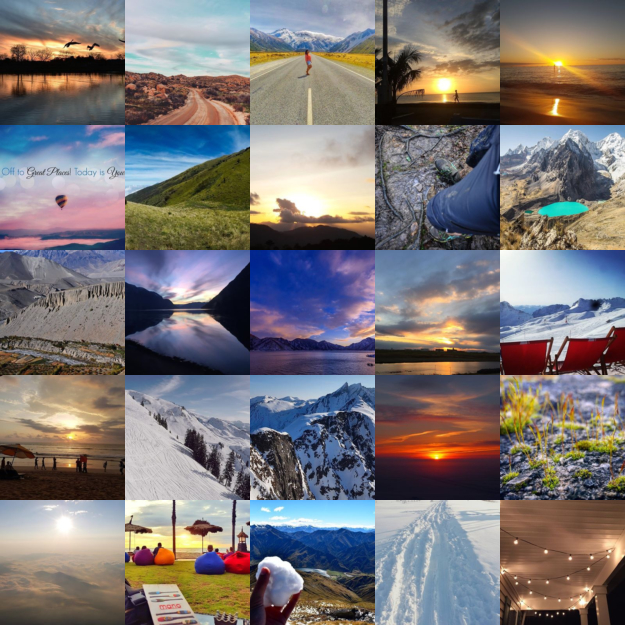}
        \caption{Low Neuroticism}
        \label{fig:max_low_n}
    \end{subfigure}
     \begin{subfigure}[b]{0.32\textwidth}
        \includegraphics[width=\textwidth]{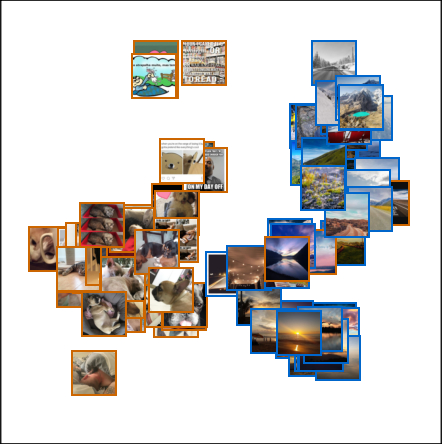}
        \caption{t-SNE visualization}
        \label{fig:tsne_neuro}
    \end{subfigure}
    \caption{\emph{Mind{P}ics} that maximally activate the Neuroticism.}
    \label{fig:max_neuro}
\end{figure*}

In our case, we inspect the output units associated to each personality trait. For example, by looking at the output of our Extraversion classifier for each of the \emph{Mind{P}ics}, we can know which ones are classified as High Extraversion with more confidence. In Figures \ref{fig:max_openness}, \ref{fig:max_cont}, \ref{fig:max_extraversion}, \ref{fig:max_agre} and \ref{fig:max_neuro} the most representative \emph{Mind{P}ics} of the High and Low scores for each trait are shown.

Among the \emph{MindPics} that maximally activate the High Openness trait we can see pictures of books, the moon and sky, while for Low Openness the most relevant pictures are love related. For High Conscientiousness most of the images are photographs of food, specially healthy food, whereas for Low Conscientiousness we mostly see pictures of people. In Extraversion it can be seen a clearly distinction between the images that maximally activate the High and Low outputs. The High Extraversion output is mostly activated by pictures of a lot of people, whereas the Low Extraversion output reacts to cats, books, and knitting images. In High Agreeablenes we mostly see flower pictures, whereas the Low score responds to pictures with text and naked torsos. Lastly, in the Neuroticism trait we observe that the High score is maximally activated by pets, whereas for Low score we see pictures of landscapes and sunsets.

\begin{figure*}[t]
    \centering
    \begin{subfigure}[b]{0.49\textwidth}
        \includegraphics[width=\textwidth]{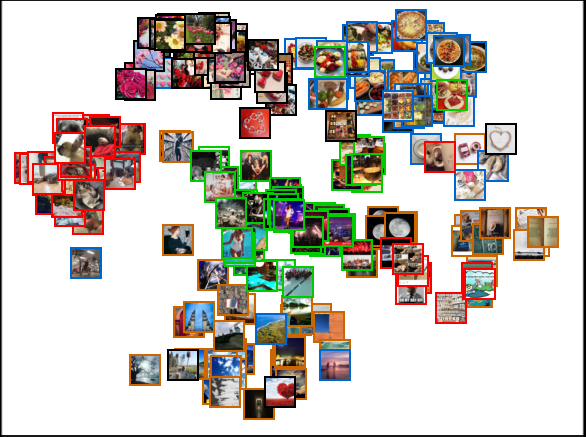}
        \caption{High correlation}
        \label{fig:tsne_all_high}
    \end{subfigure}
    ~ 
    \begin{subfigure}[b]{0.49\textwidth}
        \includegraphics[width=\textwidth]{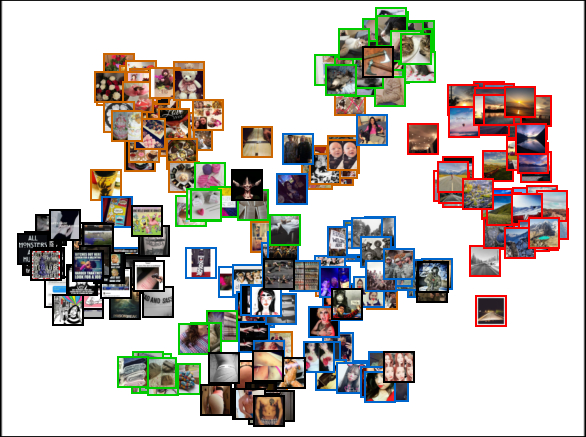}
        \caption{Lows correlation}
        \label{fig:tsne_all_low}
    \end{subfigure}
    \caption{t-SNE projection of the \emph{MindPics} that maximally activate the high (left) and low (right) scores of each trait. The colored border of the images indicates their personality trait: Openness images denoted by orange, Conscientiousness by blue, Extraversion by green, Agreeableness by black and Neuroticism by red.	}
    \label{fig:tsne_all}
\end{figure*}

In order to further analyze the behavior of our model, we study the image representations that the model produces. These representations are obtained by extracting the activations of the last feature detector of our model (the 50-layer ResNet), a high-level feature vector that describes the input image. In our case, the last layer is a Pooling layer which produces a 2048-dimensional vector. Given that these extracted features reside in a high-dimensional manifold, we can not directly visualize the images in feature space. By projecting the high-dimensional features to a 2D space, we are able to observe the underlying structure of the images representations. 

To compute this projection we use the t-Distributed Stochastic Neighbor Embedding (t-SNE) \cite{maaten2008visualizing} technique for visualizing high-dimensional data. This method retains local structure of the data and global structure of the data, so similar images appear together in the 2D space. It works by translating similarities between data to joint probabilities, and minimizes the Kullback-Leibler divergence between the probability distribution over high-dimensional points and the probability distribution over the low-dimensional data points. 

In Figure \ref{fig:tsne_all}, we have projected together the 50 images that maximally activate the High and Low scores of each trait, shown in Figures \ref{fig:tsne_open}, \ref{fig:tsne_con}, \ref{fig:tsne_extr}, \ref{fig:tsne_agre} and \ref{fig:tsne_neuro}. In the t-SNE figures we show with an orange border the pictures belonging to the High score and with a blue border the images belonging to the Low score of each trait. For the t-SNE data visualization we have used a perplexity value of $30$ and a random initialization of the embedding. The visualization of the image features projected onto a 2D plane shows that for all traits, the pictures belonging to the High and Low scores are in general well separated, with only a few errors. We must remark that t-SNE does not use the labels of the images in the projection, so these clusters arise from the image representations made by the model. 

Besides these intra-class separation, we also observe some inter-class clusters within the projected images. For example, in the Openness t-SNE projection, Figure \ref{fig:tsne_open}, we can see three different types of images of High Openness. In the right hand of the figure, we see a cluster of images of books and writings, in the middle a small cluster of pictures of the moon and in the left cluster we observe more landscapes and sky images. We also project the \emph{MindPics} that maximally activate each trait together in Figure \ref{fig:tsne_all}. We can see how even that the model is not explicitly trained to distinguish images of different traits, the representations are discriminative enough so that images of the same trait cluster together in the feature space.

\begin{figure*}[!t]
    \centering
    \begin{subfigure}[b]{0.49\textwidth}
        \includegraphics[width=\textwidth]{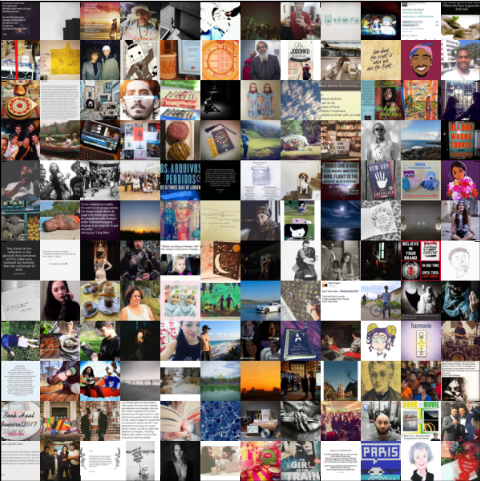}
        \caption{Images from the High Openness class}
        \label{fig:rand_high_o}
    \end{subfigure}
    ~ 
    \begin{subfigure}[b]{0.49\textwidth}
        \includegraphics[width=\textwidth]{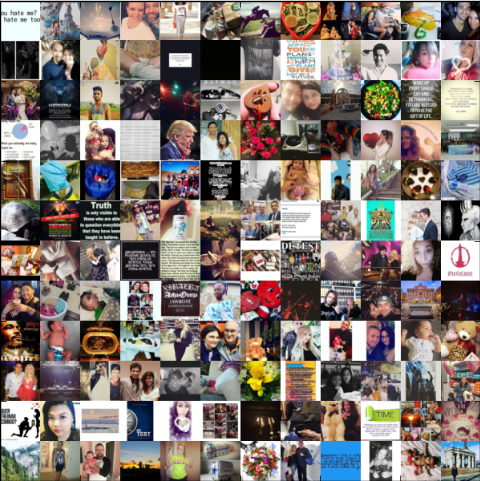}
        \caption{Images from the Low Openness class}
        \label{fig:rand_low_o}
    \end{subfigure}
    \caption{Random samples of the Openness trait.}
    \label{fig:rand_openness}
\end{figure*}

\section{Discussion}
Social media is the product of the expression of the users by means of text, images, and speech. This is a great opportunity for companies and researchers not only to know about the content of the images that are being shared, but to know about the users that create and interact with those images themselves. One interesting trait to learn from the users is personality. In fact, different methods have already been proposed to extract the users' personality from social media text and favorited images. This indicates that personality might be an underlying factor in the distribution of the users' data. In this work we made a step towards confirming this hypothesis, directly focusing on the  authors of the pictures and showing that personality remains invariant when moving from the text to the image domain. 

In particular, we showed that given the images uploaded by those users who used words correlated to the different personality traits \cite{yarkoni2010personality}, it is possible to retrieve their personality from the images. Thus, images and text are correlated, which can be explained if both depend on a third variable, which is personality. In this study, we do not recover the full spectrum of personality traits of one user, but we infer a specific personality trait that a single image conveys. The underlying hypothesis is that when a user posts a picture in a social network, the picture is not expressing everything the author has in mind, only the specific message intended by the author. In the same way, a picture does not describe the whole personality of the user, but a portion of it. So, in order to get an estimation of the whole personality profile of the users, one could analyze all the different images posted by them, because each image conveys only partial information of their personalities.

Quantitative and qualitative results were provided, showing that a neural network is indeed able to successfully learn to map \emph{Mind{P}ics} to personality traits. Moreover, the images retrieved with our model directly reflect the description of the Big Five traits, for instance, highly conscientiousness, which is described as: "Conscientiousness is a tendency to display self-discipline, act dutifully, and strive for achievement against measures or outside expectations..." in \cite{digman1990personality,barrick1991big,goldberg1990alternative}, highly correlates to images of vegetables, and people exercising. Another example of our findings is that we have seen how images of groups of people, concerts, and social settings like a bar or a restaurant are associated with high confidence to high extraversion, in contrast to pictures related to activities that one tend to do alone like reading or knitting, that are associated with low extraversion.

Lastly, in Figure \ref{fig:rand_openness} we show some random samples of the Openness trait obtained with the procedure described in this paper, the left side showing high scoring images and on the right side low scoring ones. As it can be seen, images of the same class present a lot of variability, for example \textit{High Openness} contains images of different objects like books, faces, text, drawings, and landscapes among others. This is due to the fact that some tags indeed correspond to quite abstract concepts, that leads to very different content in those related images. 

Besides the challenge of having this huge inter-class variability, we found that even images from different classes can be visually similar. For example, in Figure \ref{fig:rand_openness} we can see pictures of text and people in both \textit{High} and \textit{Low} scores of Openness. This is caused by the huge variability in real world images, and by the fact there are a lot of different images that are tagged with the same word.

\section{Conclusion}
A new framework for user personality inference has been proposed. Differently from previous approaches, the proposed framework was used to directly infer the personality of the authors of social media images retrieved according to the tags proposed in \cite{yarkoni2010personality}, which have been shown highly correlated to each of the Big Five personality traits. We then showed that it is possible to model such personality traits assigned to the retrieved images, using different strategies based on deep learning. Quantitative and qualitative results are shown, suggesting that personality is an underlying factor of the social media data distribution of users. These results open new directions of research for improving the proposed personality model by considering more words and specific images, evaluated with the supervision of psychology experts.

\vspace{6pt} 

\section*{Acknowledgements}
Authors acknowledge the funding received by the European Union's H2020 SME Instrument project under grant agreement 728633, the Spanish project TIN2015-65464-R (MINECO/FEDER), the 2016FI\_B 01163 and the 2017-SGR-1669 grants by the CERCA Programme/Generalitat de Catalunya, and the COST Action IC1307 iV\&L Net (European Network on Integrating Vision and Language) supported by COST (European Cooperation in Science and Technology). Authors would like to specially thank Ms. Daniela Rochelle Kent and Mr. Alvaro Granados Villodre for their invaluable help with the ontology of words and the images used in the experiments, respectively. We lastly gratefully acknowledge the support of NVIDIA Corporation with the donation of a Tesla K40 GPU and a GTX TITAN GPU, used for this research.

\bibliography{references}

\vspace{6pt} 

\end{document}